\documentclass[aps,prd,preprint,groupedaddress,nofootinbib]{revtex4-1}
\usepackage{bm,graphicx, wrapfig}
\usepackage{amsmath,amssymb,latexsym}
\newcommand\beq{\begin{equation}}
\newcommand\beqa{\begin{eqnarray}}
\newcommand\beqan{\begin{eqnarray*}}
\newcommand\eeq{\end{equation}}
\newcommand\eeqa{\end{eqnarray}}
\newcommand\eeqan{\end{eqnarray*}}

\begin{document}
\title{Bending angle of light in equatorial plane of Kerr-Sen Black Hole}

\author{Rashmi Uniyal}
\affiliation{Department of Physics, Government Degree College, Narendranagar 249 175, Tehri Garhwal, Uttarakhand, India}
\email{uniyal@associates.iucaa.in, rashmiuniyal001@gmail.com}

\author{Hemwati Nandan}
\affiliation{Department of Physics, Gurukula Kangri Vishwavidyalaya,
Haridwar 249 404, Uttarakhand, India}
\email{hnandan@associates.iucaa.in}

\author{Philippe Jetzer}
\affiliation{Physik-Institut, 
University of Z$\ddot{u}$rich ,
Winterthurerstrasse 190 8057 Z$\ddot{u}$rich, 
Switzerland }
\email{jetzer@physik.uzh.ch}

\begin{abstract}
\noindent
We study the gravitational lensing by a Kerr-Sen Black Hole arising in heterotic string theory.
A closed form expression for the bending angle of light in equatorial plane of Kerr-Sen Black Hole is derived as a function of impact parameter, spin and charge of the Black Hole. Results obtained are also compared with the corresponding cases of Kerr Black Hole in general relativity. 
It is observed that charge parameter behaves qualitatively similar as the spin parameter for photons travelling in direct orbits while behaves differently for photons in retrograde orbits around Black Hole. As the numerical value of the Black Hole charge increases, bending angle becomes larger in strong field limit. 
Further it is observed that this effect is more pronounced in case of direct orbits in comparison to the retro orbits. 
For both the direct and retro motion, the bending angle exceeds $2\pi$, which in turn results in multiple loops and formation of relativistic images.  
\end{abstract}

\maketitle
\section{Introduction}
\noindent
One of the consequences of Einstein's general relativity (GR) is that the light rays passing a massive body are deflected by the virtue of gravity and the resulting phenomenon is known as Gravitational lensing (GL) \citep{Schneider1992} as first observed by Eddington during the solar eclipse of 1919.  The GL, theory and observations, is one of the most important areas in modern astronomy \citep{Refsdal} and it also provides a clean and unique probe of the dark matter at all the distance scales since it is independent of the nature and physical state of the lensing mass \citep{Jetzer00}.\\ 
\noindent In fact, the existence of most compact and extreme objects in our universe such as Black Holes (BHs) and neutron stars is now well studied in view of different independent astrophysical observations. The BHs are indeed the most fascinating objects predicted by GR \citep{Chandrasekhar1983} and in addition to the BHs in GR, there are other such Black Hole (BH) solutions in various alternative theories of gravity viz. scalar-tensor theory \citep{Fujii2003}, string theory \citep{Tong2009}, braneworld scenario\citep{Sengupta2008} and loop quantum gravity \citep{Thiemann2003}. In particular, most of the BHs emerging in string theory \citep{Mignemi1993, Gibbons1988} which unifies the gravity with other three fundamental forces in nature are characterized by one or more charges associated with Yang-Mills fields. Such stringy BHs may therefore provide much deeper insight into the various properties of BH spacetimes \citep{Mignemi1993, Gibbons1988} than those of GR. The GL by a Schwarzschild BH and a Kerr BH (KBH) in the strong field limit is presented respectively in \citep{Bozza2001} and \citep{Bozza2005} in greater detail by restricting the observers in the equatorial plane. An explicit spin-dependent expression for the deflection angle in the equatorial plane of KBH is also presented in with a comparison for the case of the zero-spin BH i.e. the Schwarzschild BH in GR \citep{Iyer2009}. The detailed theoretical aspects of GL by spherically symmetric BHs in view of the perspectives for realistic observations are reviewed in \citep{Bozza2010}.\\
\noindent More recently, the closed form expression for the deflection angle of light due to a KBH is studied with a new method under the class of asymptotic approximants \citep{Barlow}. This method has been successful in the description of various physical processes like thermodynamic phase behaviour \citep{Barlow2012, Barlow2014, Barlow2015} and the solution of nonlinear boundary value problems \citep{Barlow2017}.  The GL in the Kerr-Sen BH (KSBH) which arises in the low energy limit of string theory \citep{Sen1992} as a dilaton-axion generalization of the well-known KBH in GR is also performed in weak as well as strong field limits \citep{Gyulchev2010, Gyulchev2007}.  The KSBH has the physical properties similar to the BHs arising in Einstein-Maxwell theory, but still those can be distinguished in several aspects \citep{Sen1992, Blaga2001}. However, a careful investigation to have a closed-form expression for the bending angle of light as a function of BH spin and charge is still needed.\\ 
\noindent The main objective of this paper is to study the GL by a Kerr-Sen BH  in equatorial plane 
to have exact closed-form solutions for the deflection angle of light such that both the strong and weak field limits are satisfied \citep{Barlow, Barlow2012, Barlow2014, Barlow2015, Barlow2017}. In the present work, we have followed the approach used in \citep{Iyer2009}, hence the results can be interpreted as the explicit generalisations of the results obtained in \citep{Iyer2009} for KBH. The main difference between our approach and the work done in \citep{Gyulchev2010, Gyulchev2007} is that we have obtained an explicit expression for the bending angle for both the cases i.e. direct and retrograde motion. The final expression for bending angle depends on BH mass and spin (i.e. angular momentum per unit mass of the BH) parameters. \\
\noindent In present article, the exact deflection angle is derived not only in terms of impact parameter as in Schwarzschild BH case rather in terms of several external parameters viz. (BH mass, spin parameters). Similar approach for the study of the effect of the presence of plasma on gravitational lensing and relativistic images formed by Schwarzschild BH is presented in \citep{T2013} and \citep{T2014} in detail. The organisation of this paper is as follows. The structure of KSBH spacetime is discussed in brief in section II and the critical parameters in obtaining the exact deflection angles for null geodesics are then calculated in section III. The closed-form expression for the deflection angle as a function of impact parameter and BH spin is derived in section IV. Finally, the results obtained are concluded in section V along with the future directions.
\section{Kerr-Sen BH Spacetime}
\noindent
The KSBH spacetime is described by the following $4$D effective action \citep{Sen1992},
\begin{eqnarray}
  S=-\int d^{4}x\sqrt{-\mathcal{G}}e^{-\Phi}\bigg(-\mathcal{R}
       +\frac{1}{12}\mathcal{H}^{2}
       -\mathcal{G}^{\mu\nu}\partial_{\mu}\Phi\partial_{\nu}\Phi
       +\frac{1}{8}\mathcal{F}^{2}\bigg),
       \label{action}
\end{eqnarray}
where $\Phi$ is the dilaton field and $\mathcal{R}$ is the scalar
curvature,
$\mathcal{F}^{2}=\mathcal{F}_{\mu\nu}\mathcal{F}^{\mu\nu}$ with the field strength
$\mathcal{F}_{\mu\nu}=\partial_{\mu}\mathcal{A}_{\nu}-\partial_{\nu}\mathcal{A}_{\mu}$
which corresponds to the Maxwell field
$\mathcal{A}_{\mu}$, and
$\mathcal{H}^{2}=\mathcal{H}_{\mu\nu\rho}\mathcal{H}^{\mu\nu\rho}$
with $\mathcal{H}_{\mu\nu\rho}$ given by
\begin{eqnarray}
  \mathcal{H}_{\mu\nu\rho}&=&\partial_{\mu}\mathcal{B}_{\nu\rho}
                 +\partial_{\nu}\mathcal{B}_{\rho\mu}
                 +\partial_{\rho}\mathcal{B}_{\mu\nu}
                 -\frac{1}{4}\bigg(\mathcal{A}_{\mu}\mathcal{F}_{\nu\rho}
                 +\mathcal{A}_{\nu}\mathcal{F}_{\rho\mu}
                 +\mathcal{A}_{\rho}\mathcal{F}_{\mu\nu}\bigg),\label{Hmunurho}
\end{eqnarray}
where the last term in Eq.(\ref{Hmunurho}) is the gauge
Chern-Simons term however $\mathcal{G}_{\mu\nu}$ as appeared in
Eq.(\ref{action}) are the covariant components of the metric in the
string frame, which are related to the Einstein metric by
$g_{\mu\nu}=e^{-\Phi}\mathcal{G}_{\mu\nu}$. 
The Einstein metric for KSBH, the
non-vanishing components of $A_\mu$, $B_{\mu\nu}$ and the
dilaton field respectively read as below \cite{Sen1992},
\begin{eqnarray}
  ds^{2}=-\bigg(\frac{\Delta-a^{2}\sin^{2}\theta}{\Sigma}\bigg)dt^{2}
         +\frac{\Sigma}{\Delta}dr^{2}
         -\frac{4\mu ar\cosh^{2}\alpha\sin^{2}\theta}{\Sigma}dtd\phi
         +\Sigma d\theta^{2}+\nonumber\\
         \hspace{1.2cm}\frac{\Xi\sin^{2}\theta}{\Sigma}d\phi^{2},
\label{metric}
\end{eqnarray}
\begin{eqnarray}
  \mathcal{A}_{t}&=&\frac{\mu r\sinh 2\alpha}{\sqrt{2}\Sigma},
  \;\;\;\mathcal{A}_{\phi}=\frac{\mu\,a\,r\sinh 2\alpha\sin^{2}\theta}{\sqrt{2}\Sigma},\\
  \mathcal{B}_{t\phi}&=&\frac{2a^{2}\mu r\sin^{2}\theta\sinh^{2}\alpha}{\Sigma},
  \;\Phi=-\frac{1}{2}\ln \frac{\Sigma}{r^{2}+a^{2}\cos^{2}\theta},
\end{eqnarray}
where the metric functions are described as,
\begin{eqnarray}
  \Delta&=&r^{2}-2\mu r+a^{2},\\
  \Sigma&=&r^{2}+a^{2}\cos^{2}\theta+2\mu r\sinh^{2}\alpha,\\
  \Xi&=&\bigg(r^{2}+2\mu r\sinh^{2}\alpha+a^{2}\bigg)^{2}
            - a^{2}\Delta\sin^{2}\theta.
\end{eqnarray}
The parameters $\mu$, $\alpha$ and $a$ are related to the physical
mass $M$, charge $Q$ and angular momentum $J$ as follows,
\begin{eqnarray}
  M=\frac{\mu}{2}(1+\cosh 2\alpha),\hspace{5mm}\;\;
  Q=\frac{\mu}{\sqrt{2}}\sinh^{2}2\alpha,\hspace{5mm}\;\;
  J=\frac{a\mu}{2}(1+\cosh 2\alpha)\hspace{0.5mm}.\label{equation}
\end{eqnarray}
Solving Eq.(\ref{equation}), one can obtain,
\begin{eqnarray}
 \sinh^{2}\alpha=\frac{Q^{2}}{2M^{2}-Q^{2}},\hspace{5mm}\;\;
 \mu=M-\frac{Q^{2}}{2M}\hspace{0.5mm}.\label{relation}
\end{eqnarray}
Then the parameters $\alpha$ and $\mu$ in the metric (\ref{metric})
can be eliminated accordingly. 
For a nonextremal BH, there exist two horizons, determined by $\Delta(r)=0$ as,
\begin{eqnarray}
 r_{\pm}=M-\frac{Q^{2}}{2M}\pm\sqrt{\bigg(M-\frac{Q^{2}}{2M}\bigg)^{2}-a^{2}}\hspace{1mm},
 \label{horizons}
\end{eqnarray}
where $r_{+}$ and $r_{-}$ represent the outer and the inner horizons of the BH respectively.
The case of extremal KSBH requires,
\begin{eqnarray}
 Q^{2}=2M(M-a)\hspace{1mm}.
\end{eqnarray}
The respective ranges of the parameters $a$ and $Q$ are bounded as below, 
\begin{eqnarray}
  0\leq a\leq M,\hspace{5mm}\;\;
  0\leq Q\leq \sqrt{2}M\hspace{1mm}\;\;.
\end{eqnarray}
Here, both the parameters $a$ and $Q$ are considered to be positive and for an extremal KSBH, the two horizons coincide with each other.
 One can easily obtain the first integral of radial and latitudinal coordinates (i.e. $r$ and $\phi$) for null geodesics as \citep{Uniyal2018},


\begin{equation}
{\dot{r}}^2\,=\,E^2\left(1+\frac{a^2}{r(r+x)}+\frac{2Ma^2}{r(r+x)^2}\right)-\frac{4Ma}{r(r+x)^2}EL+L^2\left(-\frac{1}{r(r+x)}+\frac{2M}{r(r+x)^2}\right),
\label{eq:u-radial}
\end{equation}
and
\begin{equation}
\dot{\phi}\,=\,\frac{L}{\Delta}\left(1-\frac{2M}{r+x}+\frac{2Ma}{r+x}\frac{E}{L}\right),
\label{eq:u-phi}
\end{equation}
where $x=Q^2/2M$ and $M=GM_{\star}/c^2$ (gravitational radius) with $M_{\star}$ defined as the physical mass of the BH. 
The time derivatives in Eq.(14) and Eq.(15) are taken with respect to the coordinate time variable $t$ defined as, $t=c\tau$ where $\tau$ represents the physical time.
Here one can define,
\begin{equation}
b_s\,=\,s\mid\frac{L}{E}\mid\,\equiv\,{sb},
\label{eq:b-var}
\end{equation}
\textbf{where the parameter $s=+{1}$ for direct orbits and $s=-1$ for retrograde orbits \citep{Iyer2009}.}
Using Eq.(\ref{eq:b-var}) in Eq.(\ref{eq:u-radial}), the radial velocity can be re-expressed as,
\begin{equation}
{\dot{r}}^2\,=\,L^2\left(\frac{1}{b^2}+\frac{a^2}{b^2r(r+x)}+\frac{2Ma^2}{b^2r(r+x)^2}-\frac{4Ma}{b_{s}r(r+x)^2}-\frac{1}{r(r+x)}+\frac{2M}{r(r+x)^2}\right).
\label{eq:u-radial-b}
\end{equation}
\section{Critical Parameters}
\noindent Using the change of variable as $u=1/r$, the orbit equation can be obtained from Eq.(\ref{eq:u-phi}) and Eq.(\ref{eq:u-radial-b}) as,
\begin{equation}
\left(\frac{du}{d\phi}\right)^2\,=\,\frac{\left(a^2u^2+(1-xu)(1-2Mu)\right)^2}{\left(1-2Mu+2Mau/b_s\right)^2(1-xu)}\mathcal{B}(u),
\label{eq:orbit-eq}
\end{equation}
where,
\begin{equation}
\mathcal{B}(u)\,=\,\frac{(1-xu)}{b^2}+\left(\frac{a^2}{b^2}-1\right)u^2+\left(1-\frac{a}{b_s}\right)^22Mu^3.
\label{eq:B-function}
\end{equation}
\par
\noindent We will further consider the case of one real negative root $u_1$ and two real distinct positive roots $u_2$ and $u_3$ given in terms of two intermediate constants $P$ and $Q$ that allow one to line up the order such as, $u_1<u_2<u_3$ \citep{Iyer2009},
\begin{eqnarray}
u_1\,=\,\frac{P-2M-Q}{4Mr_0},\\
u_2\,=\,\frac{1}{r_0},\\
u_3\,=\,\frac{P-2M+Q}{4Mr_0}.
\end{eqnarray}
By comparing the coefficients in $\mathcal{B}$(u) to those in the original polynomial in Eq.(\ref{eq:B-function}), one can first obtain the following relationship between $P$ and $\{a,b,s,r_0\}$,
\begin{equation}
P\,=\,{r_0}\frac{\left(1-\frac{a}{b_s}\right)}{\left(1+\frac{a}{b_s}\right)}.
\label{eq:P-var}
\end{equation}
It leads to the following relation between the critical parameters \citep{Iyer2009},
\begin{equation}
r_{sc}\,=\,3M\frac{\left(1-\frac{a}{b_s}\right)}{\left(1+\frac{a}{b_s}\right)}.
\label{eq:r-critical}
\end{equation}
Comparing the other coefficients of the cubic polynomial $\mathcal{B}(u)$ given by Eq.(\ref{eq:B-function}), one can also obtain the following additional expressions,
\begin{equation}
Q^2\,=\,(P-2M)(P+6M)\,+\,\frac{8M{r_0}^2x}{b^2\left(1-a/{b_s}\right)^2},
\end{equation}
\begin{equation}
Q^2-\left(P-2M\right)^2\,=\,\frac{8M{r_0}^3}{b^2\left(1-a/{b_s}\right)^2}.
\end{equation}
The intermediate variables $P$ and $Q$ can be eliminated by combining the above relations to obtain a simple cubic equation involving the impact parameter and the distance of closest approach as below,
\begin{equation}
{r_0}^3-x{r_{0}}^2-b^2\left(1-\frac{a^2}{{b_s}^2}\right)r_0+2Mb^2\left(1-\frac{a}{b_s}\right)^2\,=\,0,
\label{eq:cubic-r-zero}
\end{equation}
The Eq.(\ref{eq:cubic-r-zero}) can be solved to obtain the solution as,
\begin{equation}
r_0\,=\,\frac{2\Theta}{\sqrt{3}}\sqrt{\left(b^2\left(1-\frac{a^2}{b^2}\right)+\frac{x^2}{3}\right)}.
\end{equation}
where,
\begin{equation}
\Theta\,=\,\cos\left[\frac{1}{3}\arccos\bigg\lbrace-\frac{\sqrt{3}}{18}\left(\frac{54Mb^2\left(1-\frac{a}{b_s}\right)^2-9b^2x\left(1-\frac{a^2}{b^2}\right)-2x^3}{\left(b^2\left(1-\frac{a^2}{b^2}\right)+\frac{x^2}{3}\right)^{3/2}}\right)\bigg\rbrace\right].
\end{equation}
\noindent The above relation among the distance of closest approach and the invariant impact parameter is extremely important in Strong Deflection Limit (SDL) as well as Weak Deflection Limit (WDL) series expansions in terms of the invariant normalised quantity $b^{\prime}$. The expression also reduces to the corresponding Schwarzschild limits when charge and spin parameters are zero.
\par\noindent In the strong deflection limit, $P=3M$and the following expressions involving the critical quantities:
\begin{equation}
r_{sc}\,=\,3M\frac{\left(1-\frac{a}{b_{sc}}\right)}{\left(1+\frac{a}{b_{sc}}\right)}
\label{eq:r-sc}
\end{equation}
and
\begin{equation}
\left(b_{sc}+a\right)^3\,=\,27M^2\left(b_{sc}-a\right)-9Mx\left(b_{sc}+a\right).
\label{eq:b-sc}
\end{equation}
Eq.(\ref{eq:r-sc}) and Eq.(\ref{eq:b-sc}) exactly resemble with the previous results obtained for Schwarzschild and Kerr metrics \citep{Iyer2009}. Combining Eq.(\ref{eq:r-sc}) and Eq.(\ref{eq:b-sc}), leads to the following relation between the critical values of parameters,
\begin{equation}
{b_{sc}}^2\,=\,3{r_{sc}}\left(r_{sc}-x\right)+a^2.
\end{equation}
In order to solve the cubic Eq.(\ref{eq:b-sc}), one needs to consider the direct and retrograde motion separately (see Figs. 1 and 2).
\begin{figure}[h!]
\includegraphics[width=8cm, height=6cm]{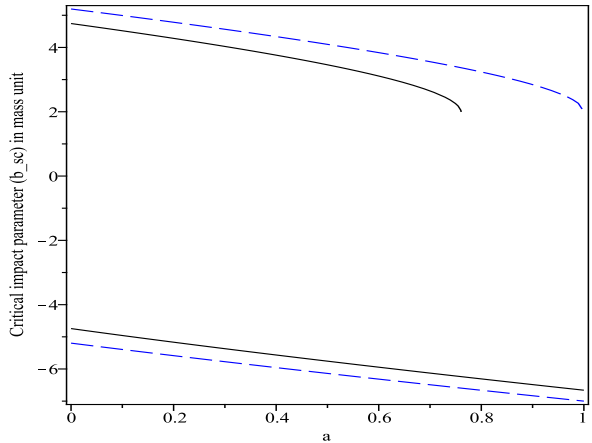}
\caption{Here solid and dashed lines represent the critical impact parameter (in mass unit) for KSBH and KBH respectively with $x=0.5$ for KSBH; further upper portion of the plot corresponds to $b_{+c}$ solutions while the lower portion of the plot corresponds to the $b_{-c}$ solutions.}
\end{figure}
\begin{figure}[h!]
\includegraphics[width=10cm, height=8cm]{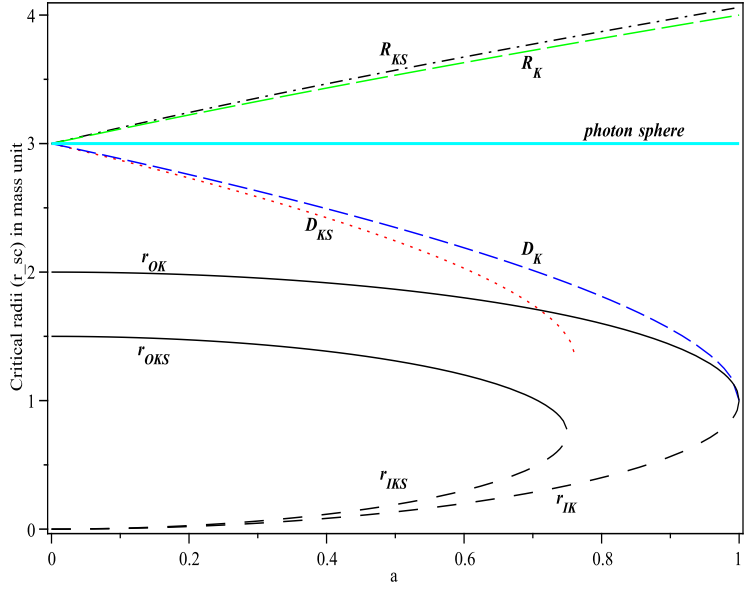}
\caption{Variation of critical radii in mass units with spin parameter $a$. Here $D_K$ and $D_{KS}$ depict the corresponding radii for direct orbits and $R_K$ and $R_{KS}$ depict the corresponding radii for retro orbits of KBH and KSBH respectively. Lower portion of the plot represents inner and outer horizons for KBH and KSBH. }
\end{figure}
\newpage 
\noindent For direct orbits (i.e. $s=+1$), Eq.(\ref{eq:b-sc}) reduces to,
\begin{equation}
\left(b_{+c}+a\right)^3-\left(27M^2-9Mx\right)\left(b_{+c}+a\right)+54aM^2\,=\,0,
\end{equation}
and solution of the above equation is given by,
\begin{equation}
b_{+c}\,=\,-a+2\sqrt{3M(3M-x)}\cos\left[\frac{1}{3}\arccos\left({\frac{-3a}{3M-x}}{\sqrt{\frac{3M}{3M-x}}}\right)\right].
\end{equation}
\noindent However for direct orbits (i.e. $s=-1$), Eq.(\ref{eq:b-sc}) reduces to,
\begin{equation}
\left(b_{-c}-a\right)^3-\left(27M^2-9Mx\right)\left(b_{-c}-a\right)-54aM^2\,=\,0,
\end{equation}
and the corresponding solution reads as,
\begin{equation}
b_{-c}\,=\,-a-2\sqrt{3M(3M-x)}\cos\left[\frac{1}{3}\arccos\left({\frac{3a}{3M-x}}{\sqrt{\frac{3M}{3M-x}}}\right)\right].
\end{equation}
Above two solutions can be clubbed together in the following form,
\begin{equation}
b_{sc}\,=\,-a+2s\sqrt{3M(3M-x)}\cos\left[\frac{1}{3}\arccos\left({\frac{-3sa}{3M-x}}{\sqrt{\frac{3M}{3M-x}}}\right)\right].
\label{critical-parameter}
\end{equation}
Using Eq.(\ref{eq:b-sc}), one can then obtain the critical value of the distance  of closest approach as,
\begin{equation}
r_{sc}\,=\,x+\frac{2}{3}(3M-x)\left[1+\cos\left(\frac{2}{3}\arccos\left({\frac{-3sa}{3M-x}}{\sqrt{\frac{3M}{3M-x}}}\right)\right)\right].
\label{critical-radii}
\end{equation}
\vspace{2mm}
\par\noindent The polynomials appearing in Eq.(\ref{eq:orbit-eq}) other than $\mathcal{B}(u)$ (given in Eq.(\ref{eq:B-function})) can be written in terms of partial fractions as,
\begin{equation}
\frac{1-2Mu(1-w_s)}{(1-xu)(1-2Mu)+a^2u^2}=\frac{C_{+}}{u_{+}-u}+\frac{C_{-}}{u_{-}-u},
\label{eq:partial-fraction}
\end{equation}
where $w_s=a/b_s$ and $u_{\pm}$ are the roots of the polynomial $(1-xu)(1-2Mu)+a^2u^2$, given as:
\begin{equation}
u_\pm\,=\,\frac{2M+x\pm\sqrt{(2M+x)^2-4(2Mx+a^2)}}{2(2Mx+a^2)}.
\end{equation}
Solving Eq.(\ref{eq:partial-fraction}) for $C_{+}$ and $C_{-}$, one obtains:
\begin{equation}
C_{+}\,=\,\frac{M(1-w_s)(2M+x+\sqrt{(2M+x)^2-4(2Mx+a^2)}-(2Mx+a^2)}{2a^2\sqrt{(2M-x-2a)(2M-x+2a)}},
\end{equation} 
and
\begin{equation}
C_{-}\,=\,\frac{(2Mx+a^2)-M(1-w_s)(2M+x+\sqrt{(2M+x)^2-4(2Mx+a^2)}}{2a^2\sqrt{(2M-x-2a)(2M-x+2a)}}.
\end{equation}
\section{Bending Angle for Light rays}
\noindent Now if one considers a light ray starting in an asymptotic region and approaching a BH, with $r_0$ as distance of its closest approach. An emerging light ray reaches upto an observer in asymptotic region. 
 Now one can easily express the involved integrals in terms of elliptical integrals of third kind to obtain the exact expressions for the bending angle \citep{Iyer2009},
 \begin{widetext}
\beqa
\alpha &=& -\pi+\;\sqrt{\frac{2}{M}}\;\;
\frac{C_+}{1-\omega_s}\int_0^{1/r_0} \frac{du}{(u_+ -u)
\sqrt{(u-u_1)(u-u_2)(u-u_3)}} \nonumber \\
\nonumber\\
&&\qquad \qquad\qquad+\; \sqrt{\frac{2}{M}}\;\;\frac{C_-}{1-\omega_s}
\int_0^{1/r_0} \frac{du}{(u_- -u)\sqrt{(u-u_1)(u-u_2)(u-u_3)}}\nonumber\\
&=&-\pi+\;\sqrt{\frac{2}{M}}\;
\frac{C_+}{1-\omega_s}\left[\int_{u_1}^{u_2}\frac{du}
{(u_+ -u)\sqrt{(u-u_1)(u_2-u)(u_3-u)}} \right. \nonumber \\
&&\qquad\qquad\qquad\qquad\qquad\qquad-\left. \int_{u_1}^0 \frac{du}{(u_+ -u)
\sqrt{(u-u_1)(u_2-u)(u_3-u)}}\right] \nonumber \\ 
&&\qquad+\;\sqrt{\frac{2}{M}}\;\frac{C_-}{1-\omega_s} 
\left[ \int_{u_1}^{u_2}\frac{du}{(u_- -u)
\sqrt{(u-u_1)(u_2-u)(u_3-u)}}\right. \nonumber \\
&&\qquad\qquad\qquad\qquad\qquad\qquad\qquad-\left. \int_{u_1}^0 \frac{du}{(u_- -u)
\sqrt{(u-u_1)(u_2-u)(u_3-u)}}\right]\nonumber
\eeqa

\beq
\label{ExactKAlpha}
\alpha=-\pi+\frac{4}{1-\omega_s}\sqrt{\frac{r_0}{Q}} 
\bigg\lbrace\Omega_+ \Bigl[\Pi(n_+,k)
-\Pi(n_+,\psi,k)\Bigr]+\Omega_- \Bigl[\Pi(n_-,k)-\Pi(n_-,\psi,k)
\Bigr]\bigg\rbrace. 
\eeq
\end{widetext}
where $\Pi(n_\pm,k)$ and $\Pi(n_\pm,\psi,k)$ 
are the complete and the incomplete elliptic integrals 
of the third kind respectively (see Appendix A).  The argument $k^2$ is defined through the elliptic
integral as usual in the range $0\le k^2 \le 1$. 
The other variables in the above expression are defined
as follows:
\beqa
\Omega_\pm&=&\frac{C_\pm}{u_\pm-u_1}\nonumber \\
k^2&=&\frac{Q-P+6M}{2Q} \nonumber \\
\psi&=& \arcsin{\sqrt{\frac{Q+2M-P}{Q+6M-P}}} \nonumber \\
n_\pm&=&\frac{u_2-u_1}{u_\pm -u_1} \nonumber
\eeqa

\begin{figure}[h!]
\includegraphics[width=7.5cm, height=7.0cm]{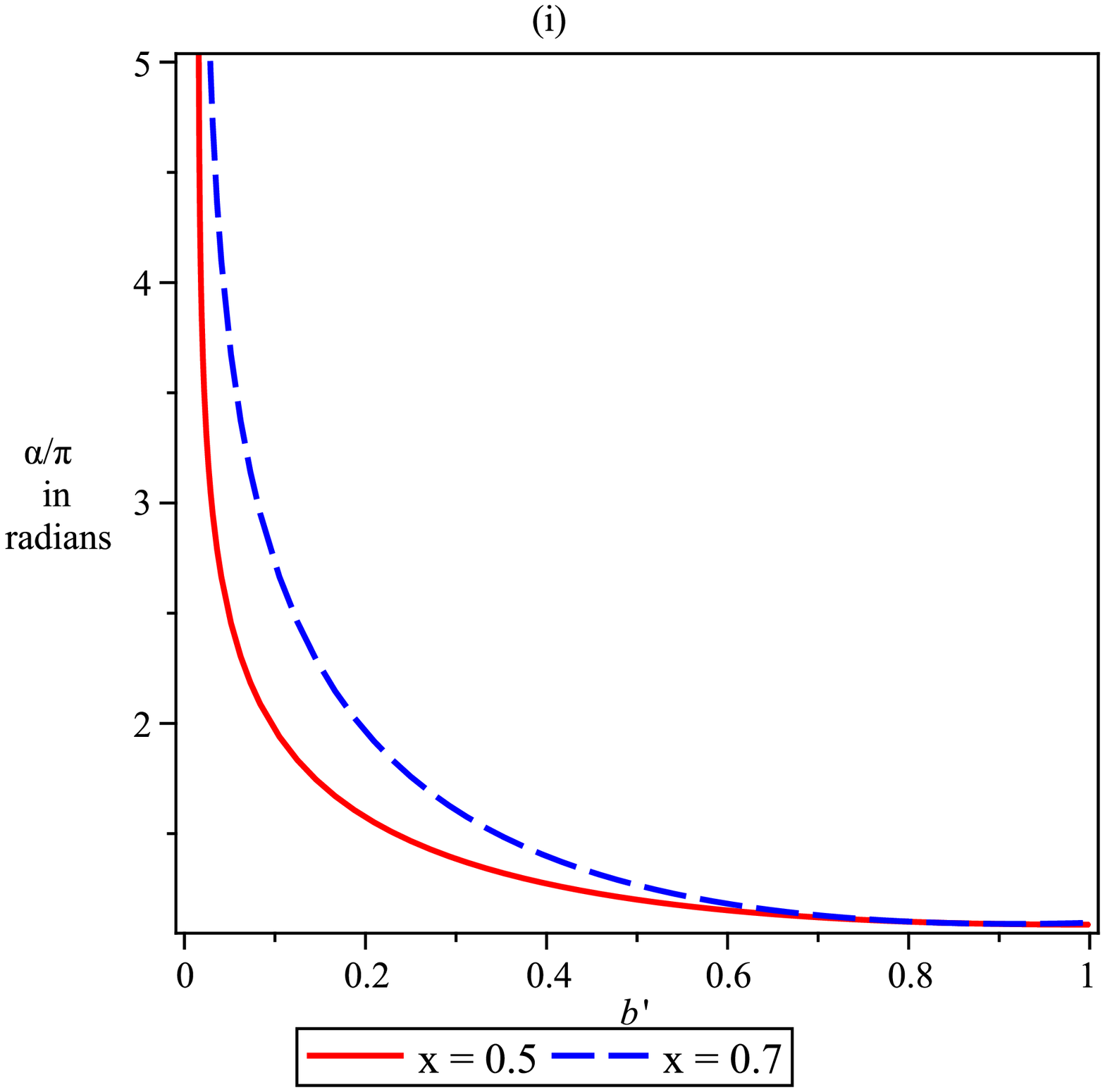}
\includegraphics[width=7.5cm, height=7.0cm]{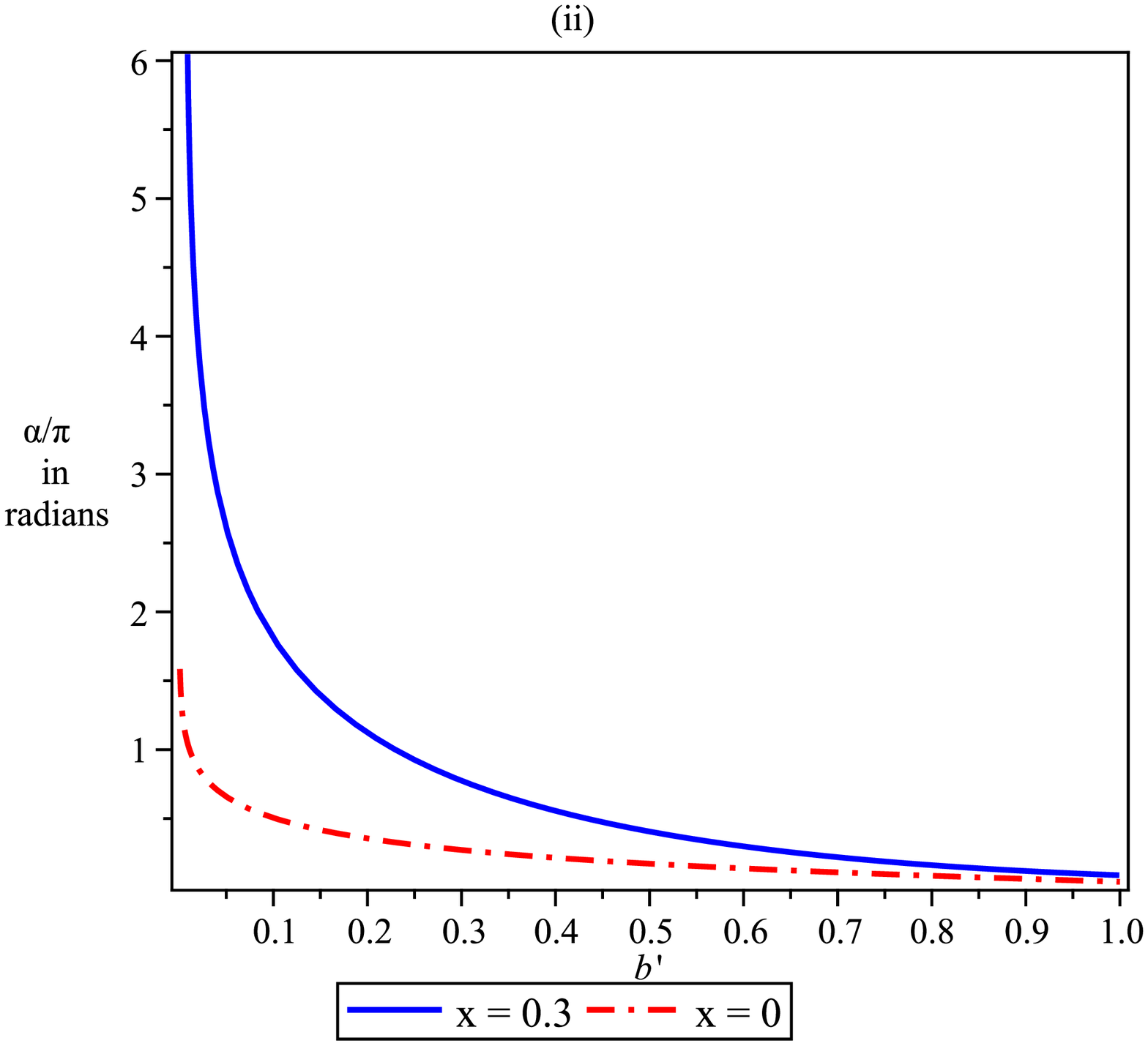}
\includegraphics[width=7.5cm, height=7.0cm]{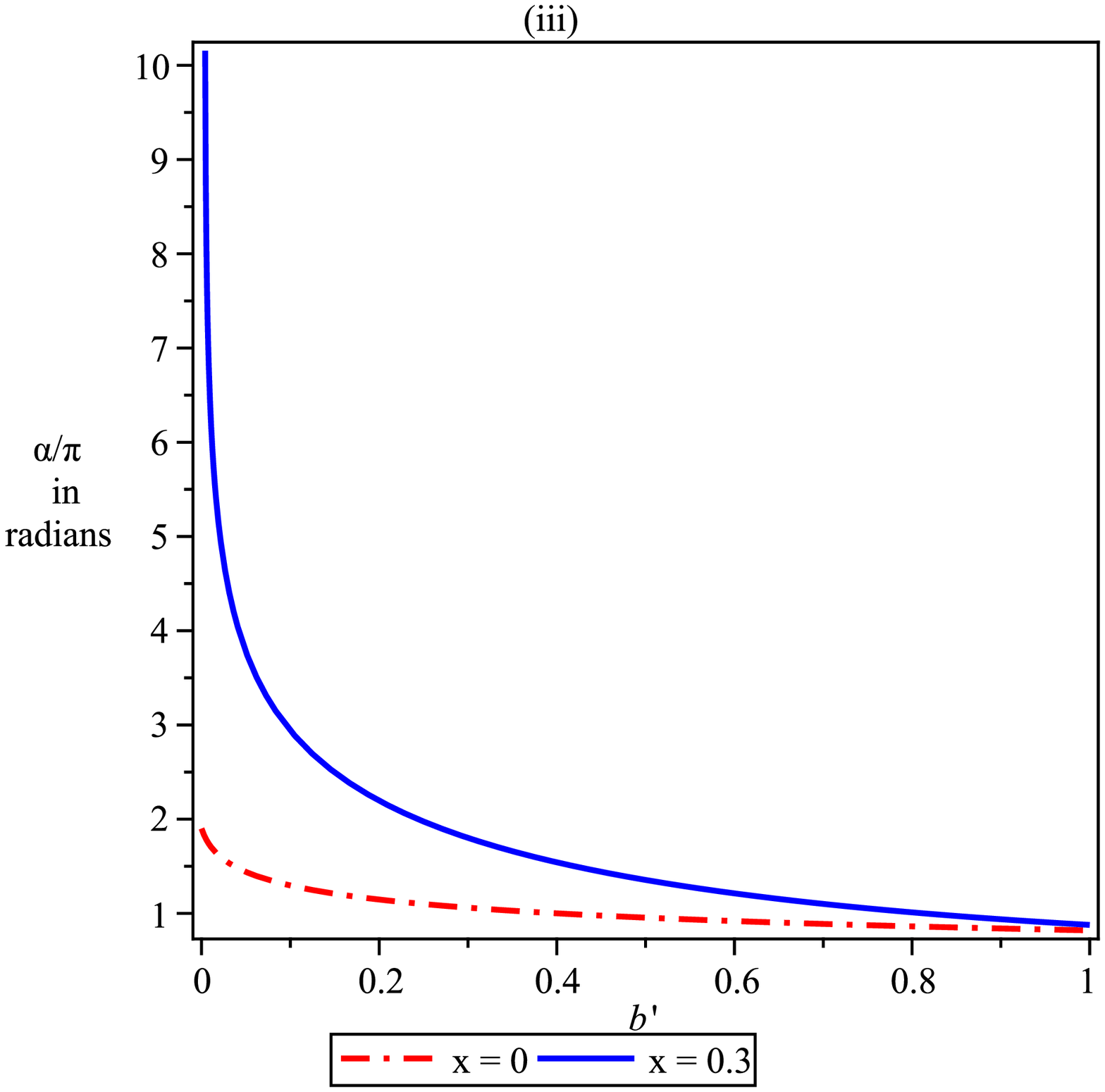}
\includegraphics[width=7.5cm, height=7.0cm]{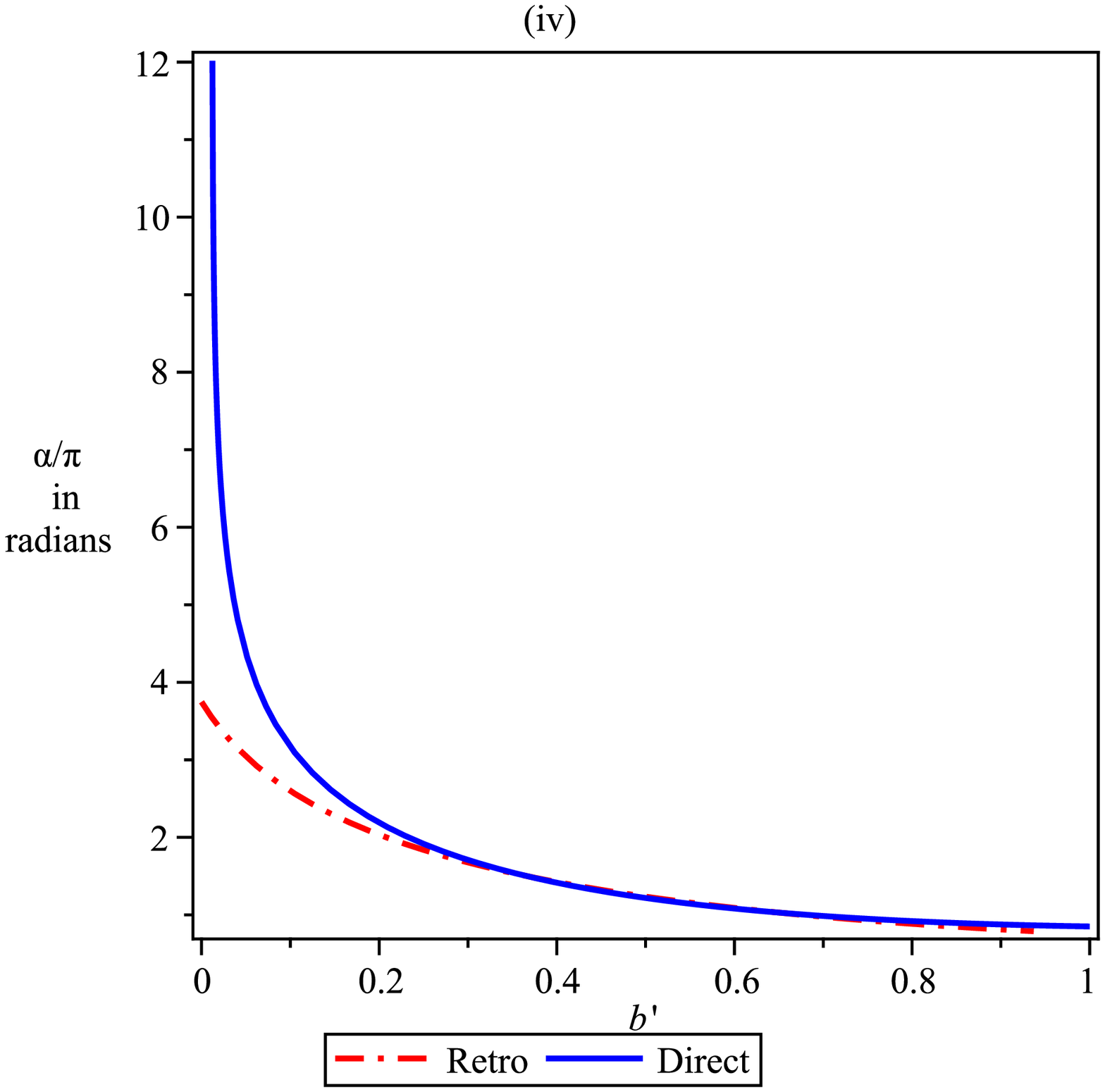}
\label{fig:bending-angle-iv}
\caption{Exact deflection angle as a function of normalised impact parameter with spin parameter value $a=0.5$.
Left section of the above plots where $b^{\prime}\rightarrow{0}$ corresponds to the strong deflection limit while the right section where $b^{\prime}\rightarrow{1}$ corresponds to weak deflection limit.}
\label{fig:bending-angle}
\end{figure}

\noindent Further the following convenient notations will be used,
\beq
h=\frac{M}{r_0}\qquad\omega_s=\frac{a}{b_s} \qquad
{\rm and}\qquad\omega_0=\frac{a^2}{M^2},
\eeq
with $\omega_s$ taking on the appropriate sign for direct and 
retrograde orbit.  
Further, we define critical 
parameters analogous to the Schwarzschild and Kerr cases in \cite{Iyer2009}:
\beq
h_{sc}=\frac{1+\omega_s}{1-\omega_s}  \qquad
{\rm and}\qquad r_{sc}=\frac{3M}{h_{sc}}.
\eeq
We also define the variable
\beq
h'=1-\frac{3h}{h_{sc}}\equiv1-3\left(\frac{M}{r_0}\right)\left(\frac{1-\omega_s}{1+\omega_s}\right).
\eeq
\noindent From a lensing perspective, we are interested in impact 
parameters beyond the critical value (SDL) extending all the way to infinity (WDL). Here the dimensionless 
quantity $b'$ is now defined as, 
\beq 
b'=1-\frac{s b_{sc}}{b},
\eeq
where the insertion of the quantity $s$ guarantees that the $b'$ stays between $0$ and $1$. 
Now, some of the intermediate variables can be eliminated to rewrite all quantities in terms of 
$h, h_{sc},\omega_0$ and $\omega_s$ as follows,
\beq
\label{bigcoeff}
\frac{r_0}{Q}=\frac{1}{h_{sc}\sqrt{\left(1-\dfrac{2h}{h_{sc}}\right)
\left(1+\dfrac{6h}{h_{sc}}\right)+\dfrac{8Mx^2}{b^2{h_{sc}}^2(1-{w_s})^2}}},
\eeq
\beq
k^2=\frac{\sqrt{\left(1-\dfrac{2h}{h_{sc}}\right)
\left(1+\dfrac{6h}{h_{sc}}\right)+\dfrac{8Mx^2}{b^2{h_{sc}}^2(1-{w_s})^2}}
+\dfrac{6h}{h_{sc}}-1}{2\sqrt{\left(1-\dfrac{2h}{h_{sc}}\right
)\left(1+\dfrac{6h}{h_{sc}}\right)+\dfrac{8Mx^2}{b^2{h_{sc}}^2(1-{w_s})^2}}},
\eeq
\beq
\psi=\arcsin{\sqrt{\frac{1-\dfrac{2h}{h_{sc}}-\sqrt{\left(1-\dfrac{2h}{h_{sc}}\right)
\left(1+\dfrac{6h}{h_{sc}}\right)+\dfrac{8Mx^2}{b^2{h_{sc}}^2(1-{w_s})^2}}}{1-\dfrac{6h}{h_{sc}}
-\sqrt{\left(1-\dfrac{2h}{h_{sc}}\right)\left(1+\dfrac{6h}{h_{sc}}\right)+\dfrac{8Mx^2}{b^2{h_{sc}}^2(1-{w_s})^2}}}}},
\eeq
\begin{widetext}
\beq
\Omega_\pm=\frac{\pm (1\pm\sqrt{1-\omega_0})
(1-\omega_s)\mp\omega_0/2}{\sqrt{1-\omega_0}\left( 1\pm\sqrt{1-\omega_0}
-\dfrac{\omega_0 h_{sc}}{4}\left[1-\dfrac{2h}{h_{sc}}-\sqrt{\left(1-\dfrac{2h}{h_{sc}}\right)
\left(1+\dfrac{6h}{h_{sc}}\right)+\dfrac{8Mx^2}{b^2{h_{sc}}^2(1-{w_s})^2}}\;\right]\right)},
\eeq
\beq
\label{nplusminus}
n_\pm=\frac{1-\dfrac{6h}{h_{sc}}-\sqrt{\left(1-\dfrac{2h}{h_{sc}}\right)
\left(1+\dfrac{6h}{h_{sc}}\right)+\dfrac{8Mx^2}{b^2{h_{sc}}^2(1-{w_s})^2}}}
{1-\dfrac{2h}{h_{sc}}-\sqrt{\left(1-\dfrac{2h}{h_{sc}}\right)
\left(1+\dfrac{6h}{h_{sc}}\right)+\dfrac{8Mx^2}{b^2{h_{sc}}^2(1-{w_s})^2}}-\dfrac{4}{\omega_0 h_{sc}}
\left(1\pm\sqrt{1-\omega_0}\right)}.
\eeq
\end{widetext}
\vspace{5mm}
\noindent
The expression of the bending angle is given by Eq.(\ref{ExactKAlpha}) after the substitution of all of the above variables. 
One may note here that the 
quantities $r_0,h, h_{sc}$, and $\omega_s$ depend on $b$, while $\omega_0=a^2/M^2$ is 
independent of $b$.  Any quantity that has an ``$s$" in the subscript takes on a negative sign 
for retro orbits. The bending angle itself stays positive since the
sign of $\phi$ in the equations of motion is determined by the incident ray in the
asymptotic region.  In other words, as the ray approaches critical on the retro side, the overall deflection
is still towards the BH even though the extent to which it is bent is smaller as compared to the static case.\vspace{5mm}\\
\noindent
In Fig.(\ref{fig:bending-angle}), the exact bending angle is plotted as a function of $b^{\prime}$ for $a=0.5$ and different values of the parameter $x$. As observed from Eq.(\ref{critical-parameter}), the critical impact parameter depends not only on the spin and charge parameters of the BH but also on the direct or retrograde motion of the photon around it. It can also be confirmed from the plots as well. As the numerical value of the BH charge increases, the bending angle of photons increases for both retro as well as direct orbits. 
Fig.[\ref{fig:bending-angle} (i, ii)] show the bending angles for photons in retrograde motion, which clearly depicts the increment in the bending angle of photon with the numerical value of the parameter $x$ (i.e. $Q^2/2M$). Fig.[\ref{fig:bending-angle} (iii)] represents the bending angle for photons in direct orbits, which again increases with the increment in the numerical value of $x$.\\
\noindent Fig.[\ref{fig:bending-angle} (iv)] shows the bending angle for retro as well as direct orbiting photons around KSBH with $a=0.5$ and $x=0.4$. This figure clearly shows that though the bending angle increases irrespective of the direct or retro motion of the photon around BH but photons orbiting in retro orbits show the smaller increment in comparison to that of orbiting in the direct orbits. 
Similar to KBH, the bending angle here also exceeds $2\pi$, which will therefore result in multiple loops and formation of relativistic images\textbf{\citep{Virbhadra2000}}, as suggested by previous studies \cite{Iyer2009}.
\newpage
\section{Conclusions and Future Directions}
\noindent 
We have studied the GL for a KSBH and derived an exact expression for the bending angle of light in its equatorial plane. 
The effect of frame-dragging on the bending angle of photons in such cases has previously been discussed for KBH \cite{Iyer2009}.
The additional charge parameter behaves similar to the spin parameter for direct orbiting photons but oppositely for retrograde orbiting photons as the bending angle increases in either case on increasing the numerical value of the charge parameter. 
Though this increment is still much larger for direct orbiting photons. 
This difference in the bending angle can clearly be visible through the shifts of the corresponding relativistic images. 
In order to study this shift in relativistic images, one needs to study the series expansion of the bending angle in weak as well as strong deflection limits. Hence, as a further work we will derive
the series expansion of the above obtained bending angle formulas in both the strong and weak field limits.
This will allow an easier comparison with similar results obtained for other BH types. \textbf{We intend to report these results in near future.}
\section*{Acknowledgments}
\noindent \textbf{The authors are indebted to the anonymous referee for the constructive comments and suggestions which helped us to improve the presentation of this paper.}
\noindent The authors RU and HN would like to thank Science and Engineering Research Board (SERB), New Delhi for
financial support through grant no. EMR/2017/000339. 
The authors RU and HN are also thankful to IUCAA, Pune (where a part of the work was completed) for support in form of academic visits under its Associateship programme. 
RU and HN would also like to thank Prof. Philippe Jetzer for the study visit to Physik-Institut, University of Zurich, Switzerland (where a part of this work was initially carried out) and for the hospitality during the stay there.
\section*{Appendix}
\noindent In Mathematica, the built-in mathematical
function for the incomplete elliptic integral of the third kind 
${\rm EllipticPi}[{\sf n},\phi,{\sf m}]$ is defined by \citep{Math},
\beq
\int_0^\phi
\left[1 - {\sf n}\sin^2 \theta\right]^{-1}\left[1 - {\sf m} \sin^2 \theta\right]^{-1/2}d \theta \nonumber
\eeq
and the complete elliptic integral of the third kind is $ {\rm EllipticPi}[{\sf n, m}] = {\rm EllipticPi}[{\sf n},\pi/2, {\sf m}]$.

\end{document}